\documentstyle[aps,eqsecnum]{revtex}

\def \en{\end{eqnarray}}
\def \bg{\begin{eqnarray}}
\def \enm{\end{mathletters}}
\def \bgm{\begin{mathletters}}

\def \ovr{\over}

\def \Orc2{{1\over c^2}}

\def \Oc2{O(1/c^2)}

\begin{document}
\centerline{\bf A Tale of Three Equations: Breit, Eddington-Gaunt, and 
Two-Body Dirac }
\centerline  {Peter Van Alstine}
\centerline{12474 Sunny Glenn Drive, Moorpark, Ca. 90125}
\centerline  {Horace W. Crater}
\centerline{The University of Tennessee Space Institute}
\centerline{Tullahoma,Tennessee 37388 }

\begin{abstract}
\noindent
G.Breit's original paper of 1929 postulates the Breit equation as a
correction to an earlier defective equation due to Eddington and Gaunt,
containing a form of interaction suggested by Heisenberg and Pauli. We
observe that manifestly covariant electromagnetic Two-Body Dirac equations
previously obtained by us in the framework of Relativistic Constraint
Mechanics reproduce the spectral results of the Breit equation but through
an interaction structure that contains that of Eddington and Gaunt. By
repeating for our equation the analysis that Breit used to demonstrate the
superiority of his equation to that of Eddington and Gaunt, we show that the
historically unfamiliar interaction structures of Two-Body Dirac equations
(in Breit-like form) are just what is needed to correct the covariant
Eddington Gaunt equation without resorting to Breit's version of retardation.
\end{abstract}

\vspace{2cm}

\vfill\eject
\newpage

\section{ Introduction}

\setcounter{section}{1}

Three score and seven years ago, Gregory Breit extended Dirac's spin-1/2
wave equation to a system of two charged particles [1]. He formed his
equation by summing two free-particle Dirac Hamiltonians with an interaction
obtained by substituting Dirac $\vec{\alpha}$'s for velocities in the
semi-relativistic electrodynamic interaction of Darwin:.

\begin{equation}
E\Psi =\left\{ \vec{\alpha}_1\cdot \vec{p}_1+\beta _1m_1+\vec{\alpha}_2\cdot 
\vec{p}_2+\beta _2m_2-\frac \alpha r[1-\frac 12(\vec{\alpha}_1\cdot \vec{%
\alpha}_2+\vec{\alpha}_1\cdot \hat{r}\;\vec{\alpha}_2\cdot \hat{r})]\right\}
\Psi .  \label{Breit}
\end{equation}

Although successful for Breit's purpose - perturbative calculation of the
electromagnetic bound-state spectrum of multi-electron atoms, Breit's
equation turned out to be neither relativistic nor a well-defined wave
equation. Nonetheless, ever since, people have applied Breit's equation in
nuclear and particle physics in situations far from its electrodynamic
origin.

In the last 20 years, some of us [2-8] have tried to remedy this by
constructing new fully covariant multiparticle relativistic quantum
descriptions. L. Horwitz has had a hand in this effort as participant and
critic. As he and F. Rohrlich pointed out in 1981 [9], for such
descriptions, the two-body problem is essentially simpler than the generic
n-body problem - reducing (just as in non-relativistic physics) to an
effective one-body problem (most simply described in terms of effective
variables introduced earlier by I.T.Todorov [10]). For most of these 20
years the authors of this paper have concentrated their efforts on using the
special two-body equations to find sensible versions of them for
interactions actually found in nature as opposed to ``toys'' like
relativistic rotors and harmonic oscillators. We have applied them (with
surprising success) to the phenomenological calculation of the q-\={q} meson
bound state spectrum [11,12].

But, of more fundamental but related importance, we have demonstrated that
our equations reproduce the spectral consequences of the original work of
Breit but in a surprising way: whereas Breit's equation yielded these
results only through first-order perturbation theory, our ''Two-Body Dirac
equations'' yield the same results through non-perturbative solution of a
fully-relativistic quantum wave-equation [13,14]. For two relativistic
spin-one-half particles interacting electromagnetically, our equations are
given as a pair of compatible Dirac equations on a single 16-component
wave-function: 
\begin{eqnarray}
D_1\psi &=&(\pi _1\cdot \gamma _1+m_1)\psi =0  \label{TwoBodyDirac} \\
D_2\psi &=&(\pi _2\cdot \gamma _2+m_2)\psi =0\text{ ,}
\end{eqnarray}

in which 
\[
\pi _i\equiv p_i-A_i\text{ , }i=1,2 
\]
and 
\begin{eqnarray*}
A_1 &=&[1-\frac 12(G+G^{-1})]p_1+\frac 12(G-G^{-1})p_2-\frac i2(\partial
G\cdot \gamma _2)\gamma _2 \\
A_2 &=&[1-\frac 12(G+G^{-1})]p_2+\frac 12(G-G^{-1})p_1+\frac i2(\partial
G\cdot \gamma _1)\gamma _1
\end{eqnarray*}
where $G^2=(1-2{\cal A}/w)^{-1}$ . $w$ is the total c.m. energy, while the
world scalar potential ${\cal A}$ is a function of the covariant spacelike
particle separation 
\begin{equation}
x_{\perp }^\mu =x_\mu +\hat{P}^\mu (\hat{P}\cdot x)
\end{equation}
which is perpendicular to the total four-momentum, $P$. For O($\alpha ^4$)
electrodynamics, ${\cal A}={\cal A}(x_{\perp })=-\alpha /r$ in which $r=%
\sqrt{x_{\perp }^2}$. In these coupled equations, the subscript $i=1,2$
stands for the $ith$ particle so that $m_1,$ and $m_2$ are the masses of the
interacting fermions. The potentials $A_i^\mu $ in this form of the Two-Body
Dirac equations appear through minimal interaction substitutions on the
particle four-momenta. They introduce mutual interactions as though each
particle were in an external potential produced by the other. (Hence, we
refer to these forms of the Two-Body Dirac equations as the ``external
potential forms''or the ``minimal interaction forms''.) The specific forms
of the covariant spin-dependent terms in the interactions as well as the
dependence of ${\cal A}$ on $x_{\bot }$ are consequences of the necessary
compatibility of the two Dirac equations 
\begin{eqnarray}
\lbrack D_1,D_2]\psi =0.
\end{eqnarray}
(In most of our previous works, we have set out these equations and their
compatibility in terms of $S$ operators that result from the $D$'s through
multiplication by $\gamma _{5i}$'s.)

Because we were originally able to obtain their versions for world scalar
interaction by rigorously `` taking the operator square roots'' of two
Klein-Gordon equations, to us they seem the natural two-body extensions of
Dirac's one-body equation hence deserving the name ``Two-Body Dirac
equations''. For the electromagnetic interaction of Eqs.(1.2-1.3), these
equations possess a family of exact solutions for para-positronium with
correct spectrum through O($\alpha ^4$)[13]. Moreover, they have not only
been derived from quantum field theory (through a relativistic
Lippmann-Schwinger equation) as the ''quantum-mechanical transform'' of the
Bethe-Salpeter equation [15,16],but their spin-independent structures have
been obtained as well from classical relativistic field theory through
Wheeler-Feynman electrodynamics. Most recently, Jallouli and Sazdjian have
completed work begun for spinless charged particles by Todorov [10] more
than 20 years ago, by obtaining our spin-one-half electrodynamic equations
from the quantum-field-theoretic eikonal approximation [17].

For all their successes, people have been slow to adopt the Two-Body Dirac
equations, preferring to stick with methods that on the one hand
artificially truncate the fully-relativistic Bethe-Salpeter equation (and
thereby produce well-known pathologies) or on the other abuse the Breit
equation by modifying its interaction structure in hopes that its
nonperturbative problems will heal themselves. The perturbative success of
Breit's equation in electrodynamics becomes a psychological barrier to use
of the fully-relativistic nonperturbative equation.

\section{History}

The historical origins of this psychological effect are intriguing. To
investigate them, we begin by noting that because the Two-Body Dirac
equations are a compatible pair, they can be rearranged in a large number of
equivalent forms. Among these is a ``Breit-form'' which in the c.m.
rest-frame reduces to the sum of two free Dirac Hamiltonian operators plus a
covariant interaction. We set out this Breit Form in a previous volume of
Foundations of Physics (in honor of F. Rohrlich)[12]. (The special
properties that allow this equation to function as a non-singular version of
the Breit equation were recently investigated by one of us in [18].)
However, we have come to realize that the particular form of this
interaction is part of an interesting controversy in the history of two-body
equations.

We decided to find out just what prompted Breit to formulate his famous
equation. Reading his paper of 1929, we saw that his work was actually
proposed as the correction of an older ``defective'' equation called by him
``the Eddington-Gaunt equation''[19] 
\begin{equation}
E\Psi =[\vec{\alpha}_1\cdot \vec{p}_1+\beta _1m_1+\vec{\alpha}_2\cdot \vec{p}%
_2+\beta _2m_2-\frac \alpha r(1-\vec{\alpha}_1\cdot \vec{\alpha}_2)]\Psi ,
\label{EddingtonGaunt}
\end{equation}
with an interaction structure first suggested by Heisenberg and Pauli
(according to Breit). We were startled to see that this equation was one
already familiar to us as the lowest-order approximation to ours (referred
to by us in [12] as ``the familiar form for four-vector interactions without
the Darwin piece''). In fact , in ''Breit-like'' form, our equation reads
[12] 
\begin{equation}
w\Psi =[\vec{\alpha}_1\cdot \vec{p}_1+\beta _1m_1+\vec{\alpha}_2\cdot \vec{p}%
_2+\beta _2m_2+w(1-\exp [-{\cal G}(x_{\bot })(1-\vec{\alpha}_1\cdot \vec{%
\alpha}_2)])]\Psi \text{ ,}  \label{RelBreit-Like}
\end{equation}
in which ${\cal G}=\ln [G]$ .

Since our covariant equation automatically has correct retardation,
apparently we have discovered in Eq.(2.2) a (higher order) corrected form of
the (Heisenberg-Pauli) ''Eddington-Gaunt'' equation that has been ''fixed''
by a mechanism different from Breit's!

What's going on here ? In fact closer reading of Breit and Eddington reveals
a fundamental reason for all this. Breit judged the Eddington-Gaunt equation
defective because of its failure after rearrangement as an equation for the
``large components'' of the wave function to reproduce the spinless Darwin
interaction: 
\begin{equation}
-\frac \alpha r[1-\frac 12(\vec{v}_1\cdot \vec{v}_2+\vec{v}_1\cdot \hat{r}\;%
\vec{v}_2\cdot \hat{r})]
\end{equation}
and its consequent failure to yield a spectrum for He in agreement with
experiment. But we recall that the Darwin interaction [20] is
straightforwardly obtained by breaking the manifest covariance of $\int \int
J_1GJ_2=\int d\tau _1\int d\tau _2\dot{x}_1^\mu \dot{x}_{2\mu
}G[(x_1-x_2)^2] $ in Wheeler-Feynman electrodynamics [21] by expanding the
(1/2 advanced + 1/2 retarded) Green's function about the instantaneous
limit. An integration by parts in the time leads to 
\begin{equation}
-\frac \alpha r[(1-\vec{v}_1\cdot \vec{v}_2)+\frac 12\vec{v}_1\cdot (1-\hat{r%
}\;\hat{r})\cdot \vec{v}_2]
\end{equation}
with the 1/2 in the second ''retardative term'' coming from a Taylor
expansion of the Green's function [22]. By redoing the Darwin interaction in
terms of the $\vec{\alpha}$'s, Breit was able to restore the missing terms.
On the other hand, if we read Eddington's notorious paper [19](and ignore
his musings about the value of the fine-structure constant $\alpha $), we
find that he arrived at the $-\frac \alpha r(1-\vec{\alpha}_1\cdot \vec{%
\alpha}_2)$ interaction by performing the covariant four-velocity goes to $%
(1,\vec{\alpha})$ substitution on the manifestly covariant JGJ coupling
(that later became the heart of Wheeler-Feynman electrodynamics) in the form
of $\dot{x}_1^\mu \;\dot{x}_{2\mu }\;\frac \alpha r$. In fact, this
covariant form appears explicitly in our Eq.(2.2) where it is inherited (but
correctly) directly from classical field theory (if we start from the
Wheeler-Feynman approach [23]) or from quantum field theory in the form of
the Bethe-Salpeter kernel in covariant Feynman Gauge [14]. This structure is
a fundamental feature of the manifestly covariant approach! By comparing
Eddington and Gaunt's Eq.(2.1) with our Eq.(2.2), we see immediately that
where they went wrong was to stop at lowest order, thereby missing an
all-important retardative recoil term!

In detail, in Eq.(2.2), if we use the $\gamma $ matrix algebra to compute
the exponential of matrix form to all orders, we find that 
\begin{eqnarray}
&&w(1-\exp [-{\cal G}(x_{\bot })(1-\vec{\alpha}_1\cdot \vec{\alpha}_2)]) \\
&=&w\frac{\exp [-{\cal G}]}4[1(3ch[{\cal G}]+ch[3{\cal G}])+\gamma
_{51}\gamma _{52}(3sh[{\cal G}]-sh[3{\cal G}])  \nonumber \\
&&+\vec{\alpha}_1\cdot \vec{\alpha}_2(sh[{\cal G}]+sh[3{\cal G}])+\vec{\sigma%
}_1\cdot \vec{\sigma}_2(ch[{\cal G}]-ch[3{\cal G}])]\text{,}  \nonumber
\end{eqnarray}
a corrected form of Eq.(69) in [12]. If we eliminate ${\cal G}$ in terms of
the potential ${\cal A}$ = $-\frac \alpha r$, we find the striking result
that to all orders in the potential: 
\begin{eqnarray}
&&w(1-\exp [-{\cal G}(x_{\bot })(1-\vec{\alpha}_1\cdot \vec{\alpha}_2)]) \\
&=&{\cal A}(1-\vec{\alpha}_1\cdot \vec{\alpha}_2)-\frac{{\cal A}^2}w(1-\vec{%
\sigma}_1\cdot \vec{\sigma}_2)-(1-\gamma _{51}\gamma _{52}+
\vec{\alpha}_2-\vec{\sigma}_1\cdot \vec{\sigma}_2)
\frac{{\cal A}^3}{w^2}{1\ovr (1-2\frac{{\cal A}}w)}\text{
.}  \nonumber
\end{eqnarray}
From this, we immediately see that the perturbative dynamics (through O($%
\alpha ^4$)) will be given by 
\begin{equation}
{\cal A}(1-\vec{\alpha}_1\cdot \vec{\alpha}_2)-\frac{{\cal A}^2}w(1-\vec{%
\sigma}_1\cdot \vec{\sigma}_2).
\end{equation}
Apparently, as a lowest-order perturbation, the second term of this must be
dynamically equivalent to the term 
\[
-\frac \alpha r\frac 12\vec{\alpha}_1\cdot (1-\hat{r}\;\hat{r})\cdot \vec{%
\alpha}_2 
\]
which when added to 
\[
-\frac \alpha r(1-\vec{\alpha}_1\cdot \vec{\alpha}_2) 
\]
makes Breit's famous 
\[
-\frac \alpha r[1-\frac 12(\vec{\alpha}_1\cdot \vec{\alpha}_2+\vec{\alpha}%
_1\cdot \hat{r}\;\vec{\alpha}_2\cdot \hat{r})]\text{ .} 
\]

\section{Equivalence?}

Not only must these terms be dynamically equivalent, but somehow our
equation must have restored the content of the Darwin interaction that Breit
discovered was spoiled by the Eddington-Gaunt equation. Breit uncovered this
defect by performing a perturbative (in orders of $1/c$) reduction to an
equation on the upper-upper component of the 16-component wave function for
both his equation and the Eddington-Gaunt equation. This rearranges the
Breit equation as a set of corrections to the non-relativistic
Schr\"{o}dinger equation with Coulomb potential. Expressed in the c.m. frame
this rearrangement becomes 
\begin{eqnarray}
H\psi =w\psi
\end{eqnarray}
in which $w$ is the total c.m. energy and 
\[
H=(m_1+{\frac{\vec{p}^2}{2m_1}}-{\frac{(\vec{p}^2)^2}{8m_1^3}})+( m_2+{\frac{%
\vec{p}^2}{2m_2}}-{\frac{(\vec{p}^2)}{8m_2^3}})+ 
\]
\[
-\alpha \big([(1-{\frac{p^2}{m_1m_2}}){\frac 1r}-{\frac 1{2m_2m_2r}}\vec{p}%
\cdot (1-\hat{r}\hat{r})\cdot \vec{p}]_{ordered} 
\]
\[
-{\frac 12}({\frac 1{m_1^2}}+{\frac 1{m_2^2}})\delta (\vec{r})-{\frac 14}{%
\frac{\vec{L}}{r^3}}\cdot [({\frac 1{m_1^2}}+{\frac 2{m_1m_2}})\vec{\sigma}%
_1+({\frac 1{m_2^2}}+{\frac 2{m_1m_2}})\vec{\sigma}_2] 
\]
\begin{eqnarray}
+{\frac 1{4m_1m_2}}(-{\frac{8\pi }3}\vec{\sigma}_1\cdot \vec{\sigma}_2\delta
(\vec{r})+{\frac{\vec{\sigma}_1\cdot \vec{\sigma}_2}{r^3}}-{\frac{3\vec{%
\sigma}_1\cdot \vec{r}\vec{\sigma}_2\cdot \vec{r}}{r^5}})\big).
\end{eqnarray}

To see what new features are contained in our truncated interaction
Eq.(2.7), we perform the same type of semirelativistic reduction on our
equation (see Appendix). We obtain a seemingly different Hamiltonian: 
\[
H=(m_1+{\frac{\vec{p}^2}{2m_1}}-{\frac{(\vec{p}^2)^2}{8m_1^3}})+( m_2+{\frac{%
\vec{p}^2}{2m_2}}-{\frac{(\vec{p}^2)}{8m_2^3}})+ 
\]
\[
-\alpha \big(-{\frac 1{2m_1m_2}}\{\vec{p}^2,1/r\}-{\frac 1{2(m_1+m_2)}}{%
\alpha /r^2}+{\frac \pi {m_1m_2}}\delta (\vec{r}) 
\]
\[
-[{\frac 14}({\frac 1{m_1^2}}+{\frac 1{m_2^2})}]({\frac{i\vec{r}\cdot \vec{p}%
}{r^3}}+\delta (\vec{r})) 
\]
\[
-{\frac 14}{\frac{\vec{L}}{r^3}}\cdot [({\frac 1{m_1^2}}+{\frac 2{m_1m_2}})%
\vec{\sigma}_1+({\frac 1{m_2^2}}+{\frac 2{m_1m_2}})\vec{\sigma}_2] 
\]
\begin{eqnarray}
+{\frac 1{4m_1m_2}}(-{\frac{8\pi }3}\vec{\sigma}_1\cdot \vec{\sigma}_2\delta
(\vec{r})+{\frac{\vec{\sigma}_1\cdot \vec{\sigma}_2}{r^3}}-{\frac{3\vec{%
\sigma}_1\cdot \vec{r}\vec{\sigma}_2\cdot \vec{r}}{r^5}})\big).
\end{eqnarray}
But, note that the $\vec{r}\cdot \vec{p}$ term and $\delta (\vec{r})$ term
in the third line of Eq.(3.3) give equivalent expectation values (both
contributing only to the ground state). Together they are equivalent to the
first term of the third line of Eq.(3.2).

Now, we come to the feature that led Breit to discard the defective
Eddington Gaunt equation in favor of his new equation: reproduction of the
correct Darwin interaction. Here this issue boils down to the equivalence
(or lack of equivalence) of the second lines of Eq.(3.3) and Eq.(3.2). As
luck would have it, Schwinger [24] has shown us just how to do this. In
order to use the virial theorem to evaluate expectation values in his
treatment of the positronium spectrum, Schwinger introduced a canonical
transformation that turns the expectation value of the second line of the
reduced Breit equation Eq.(3.2) into precisely the expectation value of the
set of terms appearing as the second line in the reduced version Eq.(3.3) of
our equation. In fact, we found some years ago that this transformation may
be used to derive the relativistic Todorov equation of electrodynamics from
classical electromagnetic field theory [25]. This takes care of the
important spin-independent Darwin interaction that served as Breit's
criterion for rejecting the Eddington Gaunt equation. But, what happened to
the intriguing spin-structure of the new term? If we look into the details
of our approximate reduction of our 16-component equation to a single
4-component equation on the upper-upper wave function (outlined in Appendix
A), we find that the spin-spin term in the equation for the upper-upper
component is exactly cancelled by a spin-spin term resulting from its
coupling to the lower-lower component. Cancellations and simplifications of
other spin dependences then yield the Fermi-Breit form of our equation -
Eq.(3.3). This establishes that the unfamiliar interaction of our Eq.(2.7)
is perturbatively quantum-mechanically equivalent to that of Breit thereby
providing a corrected form of the historical Eddington Gaunt equation. The
lesson here is that manifestly covariant techniques yield manifestly
covariant equations that produce correct spectral results if they contain
the correct interactions. We have a situation in which some of the original
investigators wrote down an interaction that was covariant (including
Eddington's realization that the interparticle separation $r$ should be
something like $\sqrt{x_{\perp }^2}$ [19]) but incomplete, while another
(Breit) replaced it with a semi-relativistic approximation to an interaction
that was correct.

\newpage

\begin{center}
{\bf Appendix A - Semirelativistic reduction of Eq.(2.2) with approximate
interaction Eq.(2.7)}
\end{center}

We write the sixteen component Dirac spinor as 
$$
\psi =\left[ \matrix{\psi_1\cr\psi_2\cr\psi_3\cr\psi_4}\right] \eqno(A.1) 
$$
in which the $\psi _i$ are four component spinors. All the matrices that
operate on this spinor are sixteen by sixteen. In the standard Dirac
representation (the subscripts on the identity 1 give the dimensionality of
the unit matrix) 
$$
\beta _1=\bigg( {%
{1_8 \atop 0}
}{%
{0 \atop -1_8}
}\bigg),\ \ \gamma _{51}=\bigg( {%
{0 \atop 1_8}
}{%
{1_8 \atop 0}
}\bigg),\ \ \beta _1\gamma _{51}\equiv \rho _1=\bigg( {%
{0 \atop -1_8}
}{%
{1_8 \atop 0}
}\bigg)\eqno(A.2) 
$$
$$
\beta _2=\bigg( {%
{\beta  \atop 0}
}{%
{0 \atop \beta }
}\bigg),\ \beta =\bigg( {%
{1_4 \atop 0}
}{%
{0 \atop -1_4}
}\bigg)\eqno(A.3) 
$$
$$
\gamma _{52}=\bigg( {%
{\gamma _5 \atop 0}
}{%
{0 \atop \gamma _5}
}\bigg),\ \gamma _5=\bigg( {%
{0 \atop 1_4}
}{%
{1_4 \atop 0}
}\bigg)\eqno(A.4) 
$$
$$
\beta _2\gamma _{52}\equiv \rho _2=\bigg( {%
{\rho  \atop 0}
}{%
{0 \atop \rho }
}\bigg),\ \rho =\bigg( {%
{0 \atop -1_4}
}{%
{1_4 \atop 0}
}\bigg)\eqno(A.5) 
$$
$$
\beta _1\gamma _{51}\gamma _{52}=\bigg( {%
{0 \atop -\gamma _5}
}{%
{\gamma _5 \atop 0}
}\bigg),\eqno(A.6) 
$$
$$
\beta _2\gamma _{52}\gamma _{51}=\bigg( {%
{0 \atop \rho }
}{%
{\rho  \atop 0}
}\bigg).\eqno(A.7) 
$$
Using $\vec{\alpha}_i=\gamma _{5i}\vec{\sigma}_i$, our truncated Breit
equation becomes 
\[
\lbrack w-m_1-m_2+{\frac \alpha r}+{\frac{\alpha ^2}{wr^2}}(1-\vec{\sigma}%
_1\cdot \vec{\sigma}_2)]\psi _1 
\]
$$
+\vec{p}\cdot \vec{\sigma}_2\psi _2-\vec{p}\cdot \vec{\sigma}_1\psi _3-{%
\frac \alpha r}\vec{\sigma}_1\cdot \vec{\sigma}_2\psi _4=0\eqno(A.8) 
$$
\[
\lbrack w-m_1+m_2+{\frac \alpha r}+{\frac{\alpha ^2}{wr^2}}(1-\vec{\sigma}%
_1\cdot \vec{\sigma}_2)]\psi _2 
\]
$$
+\vec{p}\cdot \vec{\sigma}_2\psi _1-\vec{p}\cdot \vec{\sigma}_1\psi _4-{%
\frac \alpha r}\vec{\sigma}_1\cdot \vec{\sigma}_2\psi _3=0\eqno(A.9) 
$$
\[
\lbrack w+m_1-m_2+{\frac \alpha r}+{\frac{\alpha ^2}{wr^2}}(1-\vec{\sigma}%
_1\cdot \vec{\sigma}_2)]\psi _3 
\]
$$
+\vec{p}\cdot \vec{\sigma}_2\psi _4-\vec{p}\cdot \vec{\sigma}_1\psi _1-{%
\frac \alpha r}\vec{\sigma}_1\cdot \vec{\sigma}_2\psi _2=0\eqno(A.10) 
$$
\[
\lbrack w+m_1+m_2+{\frac \alpha r}+{\frac{\alpha ^2}{wr^2}}(1-\vec{\sigma}%
_1\cdot \vec{\sigma}_2)]\psi _4 
\]
$$
+\vec{p}\cdot \vec{\sigma}_2\psi _3-\vec{p}\cdot \vec{\sigma}_1\psi _2-{%
\frac \alpha r}\vec{\sigma}_1\cdot \vec{\sigma}_2\psi _1=0\eqno(A.11) 
$$
By using Coulomb variables in which $\vec{p}\sim \alpha $ and $1/r\sim
\alpha $ we can obtain an expansion involving just the upper-upper component
wave function $\psi _1$. The expansion we desire is one through order $%
\alpha ^4$. To obtain this requires successive substitutions into Eq.(A.8)
of expressions for the lower component wave functions given in Eqs.(A.9-11)
through appropriate orders. From Eq.(A.11) we obtain 
$$
\psi _4={\frac 1{w+m_1+m_2+\alpha /r}}[(\alpha /r)\vec{\sigma}_1\cdot \vec{%
\sigma}_2-\vec{p}\cdot \vec{\sigma}_2\psi _3+\vec{p}\cdot \vec{\sigma}_1\psi
_2]\eqno(A.12) 
$$
Note that we have ignored the $\alpha ^2$ spin-dependent term in the
denominator since in the substitution into Eq.(A.8) it would have produced
terms that are higher order than $\alpha ^4$. We solve Eqs.(A.9-10) for $%
\psi _2$ and $\psi _3$ to obtain 
\[
\psi _2={\frac 1{(w+m_1-m_2+\alpha /r)(w-m_1+m_2+\alpha /r)}} 
\]
$$
\times [(w+m_1-m_2+\alpha /r)(\vec{p}\cdot \vec{\sigma}_1\psi _4-\vec{p}%
\cdot \vec{\sigma}_2\psi _1)+(\alpha /r)\vec{\sigma}_1\cdot \vec{\sigma}_2(%
\vec{p}\cdot \vec{\sigma}_1\psi _1-\vec{p}\cdot \vec{\sigma}_2\psi _4)]%
\eqno(A.13) 
$$
\[
\psi _3={\frac 1{(w+m_1-m_2+\alpha /r)(w-m_1+m_2+\alpha /r)}} 
\]
$$
\times [(w-m_1+m_2+\alpha /r)(\vec{p}\cdot \vec{\sigma}_1\psi _1-\vec{p}%
\cdot \vec{\sigma}_2\psi _4)+(\alpha /r)\vec{\sigma}_1\cdot \vec{\sigma}_2(%
\vec{p}\cdot \vec{\sigma}_1\psi _4-\vec{p}\cdot \vec{\sigma}_2\psi _1)]%
\eqno(A.14) 
$$
In Eq.(A.8) we need $\psi _4$ through order $\alpha ^2$ and thus in
Eq.(A.12) we need $\psi _2$ and $\psi _3$ only through order $\alpha $. This
in turn implies that we can drop the $\psi _4$ and interaction terms in
Eqs.(A.13-14). Performing this substitution yields the following expression
for $\psi _4$ of appropriate order for use in Eq.(A.8): 
$$
\psi _4={\frac 1{2(m_1+m_2)}}[{\frac \alpha r}\vec{\sigma}_1\cdot \vec{\sigma%
}_2-{\frac{m_1+m_2}{2m_1m_2}}\vec{p}\cdot \vec{\sigma}_1\vec{p}\cdot \vec{%
\sigma}_2]\psi _1\eqno(A.15) 
$$
For the direct $\psi _2$ and $\psi _3$ terms in Eq.(A.8) we need Eq.(A.13)
and Eq.(A.14) through order $\alpha ^3$, which in turn requires Eq.(A.15).
This leads to 
\[
\psi _2=-{\frac 1{w-m_1+m_2+\alpha /r}}[\vec{p}\cdot \vec{\sigma}_2-{\frac{%
\vec{p}\cdot \vec{\sigma}_1}{2(m_1+m_2)}(}{\frac \alpha r}\vec{\sigma}%
_1\cdot \vec{\sigma}_2-\vec{p}\cdot \vec{\sigma}_1\vec{p}\cdot \vec{\sigma}_2%
{\frac{m_1+m_2}{2m_1m_2})}]\psi _1 
\]
$$
+{\frac{(\alpha /r)}{4m_1m_2}}\vec{\sigma}_1\cdot \vec{\sigma}_2\vec{p}\cdot 
\vec{\sigma}_1\psi _1\eqno(A.16) 
$$
\[
\psi _3={\frac 1{w+m_1-m_2+\alpha /r}}[\vec{p}\cdot \vec{\sigma}_1-{\frac{%
\vec{p}\cdot \vec{\sigma}_2}{2(m_1+m_2)}(}{\frac \alpha r}\vec{\sigma}%
_1\cdot \vec{\sigma}_2-\vec{p}\cdot \vec{\sigma}_1\vec{p}\cdot \vec{\sigma}_2%
{\frac{m_1+m_2}{2m_1m_2})}]\psi _1 
\]
$$
-{\frac{(\alpha /r)}{4m_1m_2}}\vec{\sigma}_1\cdot \vec{\sigma}_2\vec{p}\cdot 
\vec{\sigma}_2]\psi _1.\eqno(A.17) 
$$
Combining all terms of Eq.(A.8) and replacing $w$ by $m_1+m_2$ in the $%
\alpha ^4$ terms we obtain 
\[
w\psi _1=[m_1+m_2-{\frac \alpha r}-\frac{\alpha ^2}{r^2(m_1+m_2)}(1-\vec{%
\sigma}_1\cdot \vec{\sigma}_2)]\psi _1 
\]
\[
+\vec{p}\cdot \vec{\sigma}_2{\frac 1{w-m_1+m_2+\alpha /r}}[\vec{p}\cdot \vec{%
\sigma}_2-{\frac{\vec{p}\cdot \vec{\sigma}_1}{2(m_1+m_2)}(}{\frac \alpha r}%
\vec{\sigma}_1\cdot \vec{\sigma}_2-\vec{p}\cdot \vec{\sigma}_1\vec{p}\cdot 
\vec{\sigma}_2{\frac{m_1+m_2}{2m_1m_2})}]\psi _1 
\]
\[
+\vec{p}\cdot \vec{\sigma}_1{\frac 1{w+m_1-m_2+\alpha /r}}[\vec{p}\cdot \vec{%
\sigma}_1-{\frac{\vec{p}\cdot \vec{\sigma}_2}{2(m_1+m_2)}(}{\frac \alpha r}%
\vec{\sigma}_1\cdot \vec{\sigma}_2-\vec{p}\cdot \vec{\sigma}_1\vec{p}\cdot 
\vec{\sigma}_2{\frac{m_1+m_2}{2m_1m_2})}]\psi _1 
\]
\[
-\vec{p}\cdot \vec{\sigma}_2{\frac{(\alpha /r)}{4m_1m_2}}\vec{\sigma}_1\cdot 
\vec{\sigma}_2\vec{p}\cdot \vec{\sigma}_1\psi _1-\vec{p}\cdot \vec{\sigma}_1{%
\frac{(\alpha /r)}{4m_1m_2}}\vec{\sigma}_1\cdot \vec{\sigma}_2\vec{p}\cdot 
\vec{\sigma}_2\psi _1 
\]
$$
+{\frac \alpha r}\vec{\sigma}_1\cdot \vec{\sigma}_2[{\frac \alpha {%
2(m_1+m_2)r}}\vec{\sigma}_1\cdot \vec{\sigma}_2-{\frac 1{4m_1m_2}}\vec{p}%
\cdot \vec{\sigma}_1\vec{p}\cdot \vec{\sigma}_2]\psi _1\eqno(A.18) 
$$
We omit most of the steps of the remaining reduction, involving powers of
Pauli matrices, commenting on just two portions of the details. The first is
that the spin-spin term in the first line cancels with the corresponding
spin-spin term in the last line that results from the identity $(\vec{\sigma}%
_1\cdot \vec{\sigma}_2)^2=3-2\vec{\sigma}_1\cdot \vec{\sigma}_2$. The second
is that the semirelativistic $\vec{p}^4$ kinetic corrections not only result
from terms at the end of the second and third lines that involve four $\vec{p%
}\cdot \vec{\sigma}_i$ factors but also from the terms at the beginning of
each of those lines, in particular 
$$
(\vec{p}\cdot \vec{\sigma}_2{\frac 1{w-m_1+m_2+\alpha /r}}\vec{p}\cdot \vec{%
\sigma}_2+\vec{p}\cdot \vec{\sigma}_1{\frac 1{w+m_1-m_2+\alpha /r}}\vec{p}%
\cdot \vec{\sigma}_1)\psi .\eqno(A.19) 
$$
One brings the denominator through and operates on $\psi _1$ using the lower
order portions of wave equation Eq.(A.18). When we perform this operation
and various other simplifications we obtain the semirelativistic reduction
given in Eq.(3.3) in the text of the paper.

\end{document}